\documentclass[prl,twocolumn,floats,superscriptaddress]{revtex4-1}
\usepackage{amsmath}
\usepackage{amssymb}
\usepackage{graphicx}
\usepackage{color}

\begin{document}

\title{Discrete synaptic events induce global oscillations in balanced neural networks}
\date{\today}

        \author{Denis S. Goldobin}
      \affiliation{Institute of Continuous Media Mechanics, Ural Branch of RAS, Acad. Korolev
street 1, 614013 Perm, Russia}
\affiliation{ Department of Theoretical Physics, Perm State University, Bukirev street 15,
614990 Perm, Russia}
\author{Matteo di Volo }
\affiliation{Universit\'e Claude Bernard Lyon 1, Institut National de la Sant\'e et de la Recherche M\'edicale, Stem Cell and Brain Research Institute U1208, Bron, France}     
                \author{Alessandro Torcini}
                \email[corresponding author: ]{alessandro.torcini@cyu.fr}
        \affiliation{Laboratoire de Physique Th\'eorique et Mod\'elisation, Universit\'e de Cergy-Pontoise,CNRS, UMR 8089, 95302 Cergy-Pontoise cedex, France}
        \affiliation{CNR - Consiglio Nazionale delle Ricerche - Istituto dei Sistemi Complessi, via Madonna del Piano 10, I-50019 Sesto Fiorentino, Italy}

\begin{abstract}
Neural dynamics is triggered by discrete synaptic inputs of finite amplitude. However, the
neural response is usually obtained within the diffusion approximation (DA) representing the
synaptic inputs as Gaussian noise. We derive a mean-field formalism encompassing synaptic shot-noise 
for sparse balanced networks of spiking neurons. For low (high) external drives (synaptic strenghts)
irregular global oscillations emerge via continuous and hysteretic transitions, correctly
predicted by our approach, but not from the DA. These oscillations display frequencies in
biologically relevant bands. 
\end{abstract}

\maketitle

\paragraph{Introduction.}

In several contexts the discrete nature of stochastic events should be taken into
account to correctly predict the system dynamics. A typical example
is represented by shot-noise, which is conveyed by pulses and is therefore discontinous,
at variance with white noise, which is associated to thermal fluctuations and is continuous \cite{sc1918}.
The inclusion of shot-noise is fundamental to fully characterize
the emergent phenomena in many fields of physics ranging from mesoscopic conductors \cite{blanter2000} 
to driven granular gases \cite{lucente2023}.

The discrete nature of the events is an innate characteristic also of the neural
dynamics, where a neuron receives inputs from other neurons via electrical pulses,
termed post-synaptic potentials (PSPs). The PSPs stimulating a neuron in the cortex
are usually assumed to be uncorrelated with small amplitudes and high arrival rates.
Therefore the synaptic inputs can be treated as a continuous Gaussian process and 
the neural dynamics can be examined at a mean-field  level within the framework of the Diffusion Approximation (DA) 
\cite{capocelli1971,tuckwell1988}. In this context, the theory of dynamical balance 
of excitation and inhibition \cite{bal1,bal2,brunel2000} represents one of the most
successfull results able to explain some of the main aspects of cortical dynamics \cite{barral2016}.

However, several experiments have shown that rare PSPs of large amplitude 
can have a fundamental impact on the cortical activity \cite{song2005,lefort2009}
and that synaptic weight distributions display a long tail towards
large amplitudes \cite{miles1990,barbour2007,buzsaki2014}.

Furthermore, networks of inhibitory neurons with low connectivity 
(in-degree $K \simeq 30-80$) have been identified in the cat visual cortex \cite{kisvarday1993}
and in the rat hippocampus \cite{sik1995} and the latter are believed to be at 
the origin of collective oscillations (COs) in the $\gamma$-band \cite{buzsaki2012}. 
Recent experiments have also shown that the cortical connections are 
definitely more sparse in primate when compared to mouse \cite{wildenberg2021}.

These experimental evidences call for the development
of a mean-field formalism able to incorporate the effect of discrete synaptic events
for diluted random networks. Population based formalisms taking into account the 
discrete nature of the synaptic events have been previously developed for 
Integrate-and-Fire models \cite{richardson2010,iyer2013, olmi2017,droste2017}.
However, such approaches are limited to stationary solutions and they 
cannot describe the emergence of oscillatory behaviours.
 
In this Letter, we introduce a {\it complete} mean-field (CMF) approach for balanced neural networks \cite{bal1},
taking into account the sparsness of the network and the discreteness of the synaptic
pulses, able to reproduce all the possible dynamical states.
 For simplicity, but without any loss of generality, we consider inhibitory balanced networks 
subject to an external excitatory drive \cite{brunel1999,bal4,wolf,matteo}.

Firstly, we illustrate that the DA cannot capture oscillatory behaviours
emerging for sufficiently low in-degree in spiking neural networks by considering conductance-
and current-based neuronal models. However, this regime is correctly reproduced by a
mean-field approach whenever the sparse and discrete synaptic inputs are taken in account. Furthermore, for Quadratic Integrate-and-Fire (QIF) 
~\cite{ermentrout1986, gutkin2022} neuronal network via the CMF approach we obtain a complete bifurcation diagram encompassing asynchronous and oscillatory regimes. 
In particular, for sufficiently low (large) excitatory drive (synaptic amplitudes) 
the CMF reveals bifurcations from the asynchronous irregular (AI) to the oscillatory irregular (OI) regime as well as a region of coexistence of these two phases not captured by the DA \cite{noi}. 
Exact event-driven simulations of large QIF networks confirm the 
sub- and super-critical Hopf bifurcations predicted within the CMF theory.
Furthermore, for low in-degrees COs in biologically relevant frequency bands 
(from  $\delta$ to $\gamma$ band) are observable  \cite{buzsaki2006}.

\paragraph{The balanced network.}

As a prototype of a dynamically balanced system we consider a sparse inhibitory network made of $N$
pulse-coupled neurons whose membrane potential evolves according to the equations
\begin{equation}\label{eq:1}
\dot{V}_{i}(t) = F(V_{i}) + I - g \sum_{j=1}^{N} \sum_{n} \epsilon_{ji} \delta(t - t_{j}^{(n)}) \enskip ;
\end{equation}
where $I$ is an external DC current, $g$ the synaptic coupling, and the last term represents
the inhibitory synaptic current. The latter is the linear superposition of instantaneous
inhibitory PSPs emitted at times $t_{j}^{(n)}$ from
the pre-synaptic neurons connected to neuron $i$.
$\epsilon_{ji}$ is the adjacency matrix of the random network with entries $1$ $(0)$
if the the connection from node $j$ to $i$ exists (or not), and
we assume the same in-degree $K =\sum_j\epsilon_{ji}$ for all neurons.
We consider two paradigmatic models
of spiking neuron: the quadratic integrate-and-fire (QIF)
with $F(V) = V^2$~\cite{ermentrout1986,wolf,laing2018,matteo,ratas2019}, which is a current-based model of
class I excitability; and the Morris-Lecar (ML) \cite{morris1981},
a conductance-based model representing a class II excitable membrane \cite{supp}.
The DC current and the synaptic coupling are assumed to scale as
$I= i_0 \sqrt{K}$ and  $g=g_0/\sqrt{K}$ as usually done in order to ensure
a self-sustained balanced state for sufficiently large in-degrees \cite{bal1,bal2,bal3,bal4,wolf,matteo}.
The times (frequencies) are reported in physical units by assuming a membrane time constant $\tau_m=10$ ms.

\paragraph{Mean-field description.}

\begin{figure}
\begin{center}
\includegraphics[width=1.0 \linewidth]{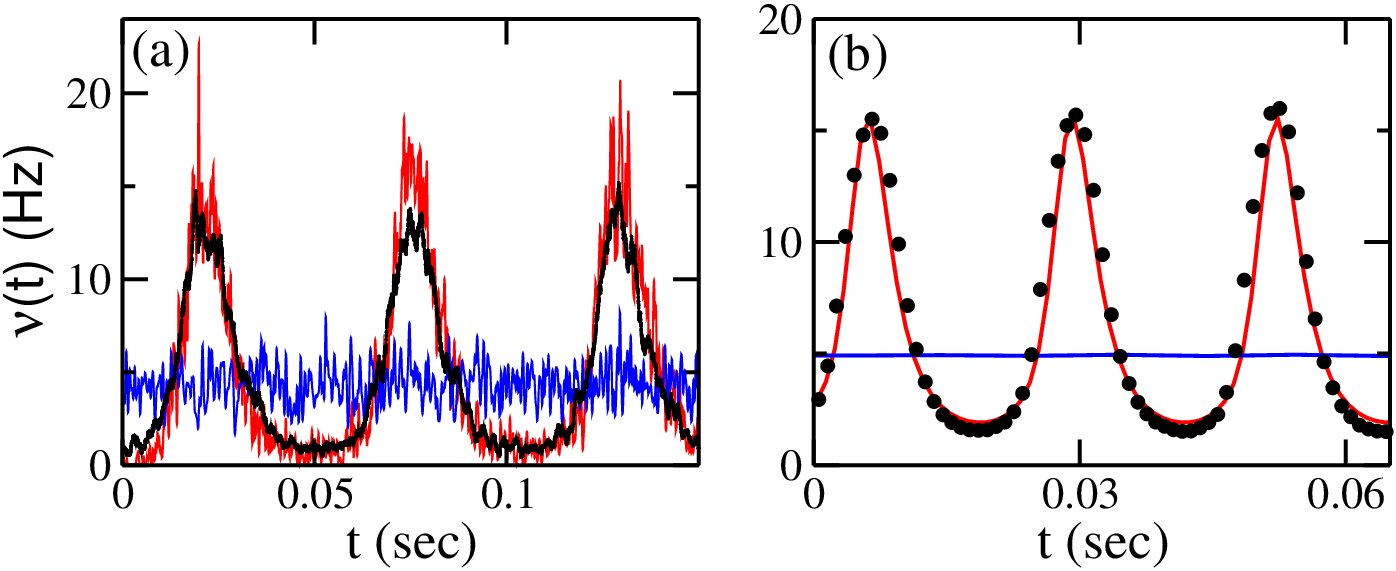}
\end{center}
\caption{Population firing rate $\nu(t)$ versus time for ML (a) and QIF (b) models:
blue (red) lines refer to diffusive (shot-noise) MF results and black circles to network simulations.
For the ML the MF shot-noise and DA results have been obtained by integrating the Langevin equation \eqref{eq:langevin},
while for the QIF by integrating \eqref{shot} and \eqref{fpe}, respectively: more details in \cite{supp}.
The parameters for the ML model are $K=20$, $i_0 = 0.1$, $g_0=5$ and network size $N=20000$, the other parameters are reported in the supplemental material \cite{supp}. For the QIF model $K=200$, $i_0 = 0.16$, $g_0=4$ and $N=80000$.
}
\label{fig1}
\end{figure}

For a sufficiently sparse network, the spike trains emitted by $K$ pre-synaptic neurons
can be assumed to be uncorrelated and Poissonian \cite{brunel1999,brunel2000}, therefore
the mean-field dynamics of a generic neuron can be represented in terms of following Langevin equation:
\begin{equation}\label{eq:langevin}
\dot{V}(t) = F(V) + I - g S(t)
\end{equation}
where $S(t)$ is a Poissonian train of $\delta$-spikes with rate $R(t) = K \nu(t)$, and $\nu(t)$ is the
population firing rate self-consistently estimated.
Usually the Poissonian spike trains are approximated within the
the DA \cite{ricciardi,tuckwell1988} as
$S(t)  =  R(t) + \sqrt{R(t)} \xi(t)$, where $\xi(t)$ is a Gaussian white noise term.
However, this approximation can fail to reproduce
fundamental aspects of the neural dynamics. Indeed, as shown in Fig. \ref{fig1} (a) for a sparse ML network,
by employing the DA in \eqref{eq:langevin} one obtains an asynchronous dynamics (blue curve), while the correct network evolution, characterized by global oscillations with frequency $f_C \simeq 18$ Hz (black dots), can be recovered only by 
explicitely taking into account the Poissonian spike trains in \eqref{eq:langevin} (red line).

In the mean-field framework the population dynamics is usually described in terms of
the membrane potential probability distribution function (PDF) $P(V,t)$,
whose time evolution is given for the QIF model by the following continuity equation
\begin{equation}
\label{shot}
{\dot P}(V,t) + \partial_V[(V^2 + I)P(V,t)] = R(t) \Delta P(V,T) 
\end{equation}
with boundary condition $\lim_{V \to \infty} V^2 P(V,t) = \nu(t)$ and
where $\Delta P(V,T) = [P(V^+,t)-P(V,t)]$ with $V^+=V + g$.
By assuming that $g$ is sufficiently small we can expand the latter term as
$
\Delta P(V,t) = \sum_{p=1}^\infty \frac{g^p}{p !} \partial_V^p P(V,t)   \enskip ;
$
and by limiting to the first two terms in this expansion we recover the
DA corresponding to the following Fokker-Planck Equation (FPE) \cite{haskell2001}
\begin{equation}
\label{fpe}
{\partial_t P(V,t)} + {\partial_V}[(V^2 + A(t))P(V,t)] =
D(t) {\partial^2_{V} P(V,t)}
\end{equation}
where $A(t) = \sqrt{K}[i_0 - g_0 \nu(t)]$ and $D(t) = g_0^2 \nu(t)/2$.
The DA can give uncorrect predictions for the QIF model, as well.
Indeed as shown in Fig. \ref{fig1} (b) the network dynamics
is oscillatory with $f_c \simeq 40$ Hz (black circles) : an evolution correctly captured by the MF equation
 \eqref{shot} (red line), while the FPE \eqref{fpe} converges to a a stable fixed point (blue curve),
Therefore to reproduce the collective dynamical regimes observable in the network
it is necessary to consider the complete continuity equation \eqref{shot}.
In this respect we have developed a CMF formalism encompassing synaptic shot-noise 
to identify the various possible regimes displayed by \eqref{shot} and to analyse their stability.

The QIF model evolution can be transformed in that of a phase oscillator,
the so-called $\theta$-neuron \cite{ermentrout1986,ermentrout2008},
by introducing the phase variable $\theta = 2 \arctan{V}$.
However, this transformation has the drawback that even uncoupled neurons
are associated to a non flat PDF of the phases, thus rendering quite difficult or even unfeasible to identify
asynchronous regimes with respect to partially synchronized ones in noisy enviroments \cite{kralemann2007,dolmatova2017}.
A more appropriate phase transformation to analyse the synchronization phenomena is the following
$\psi = 2 \arctan{ (V/\sqrt{I})} \in [-\pi,\pi]$, which leads to a uniformly rotating
phase in the absence of incoming pulses for supra-threshold neurons with $I >0$ \cite{supp}.

By considering the phase PDF $w(\psi,t)=P(V,t)\big(I+V^2\big)/(2\sqrt{I})$, Eq.~\eqref{shot} can be rewritten  
in terms of the so-called Kuramoto--Daido order parameters $z_n$ \cite{kuramoto2012, daido1992} by expanding
in Fourier space the PDF as $w(\psi,t)=(2\pi)^{-1}\sum_{n=-\infty}^{+\infty}z_ne^{-in\psi}$ with $z_0=1$ and $z_{-n}=z_n^\ast$\,. After laborious but straightforward calculations, one obtains the following evolution  equations
\begin{equation}
\dot{z}_n=i2n \sqrt{I} z_n +K\nu\left[\sum_{m=0}^{+\infty} I_{nm}(\alpha)\,z_m-z_n\right],
\label{eqFP03}
\end{equation}
where $n=1,2,3,...$\,, $\alpha\equiv g/\sqrt{I}=g_0/(\sqrt{i_0} K^{3/4})$,
\begin{align}
&I_{nm}(\alpha)\equiv\frac{1}{2\pi}\int\limits_0^{2\pi} \frac{e^{in\psi}\left(e^{-i\psi_a}\right)^m\mathrm{d}\psi}{1+\frac{\alpha^2}{2}+\alpha\sin\psi+\frac{\alpha^2}{2}\cos\psi}
\label{eqFP04}
\\
&=\!\left\{
\begin{array}{cr}
\big(\frac{\alpha}{2i-\alpha}\big)^n\;,
&m=0\,;\\[5pt]
\sum\limits_{j=1}^{\min(n,m)}
\frac{4(-1)^j(n+m-j)!\cdot\alpha^{m+n-2j}(4+\alpha^2)^{j-1}} {m(j-1)!\cdot(m-j)!\cdot(n-j)!\cdot(2i-\alpha)^{m+n}},
&m\ge 1\,.
\end{array}
\right.
\nonumber
\end{align}

The firing rate can be self-consistently determined by the flux at the firing threshold
$\lim_{V \to \infty}V^2 P(V,t)=2\sqrt{I} w(\pi,t)$, as follows
\begin{align}
\nu= 2\sqrt{I} w(\pi,t) = \frac{\sqrt{I}}{\pi}\mathrm{Re}\left(1+2\sum_{k=1}^\infty (-1)^k z_k\right).
\label{eqFP05}
\end{align}

\begin{figure}
\begin{center}
\includegraphics[width=0.95 \linewidth]{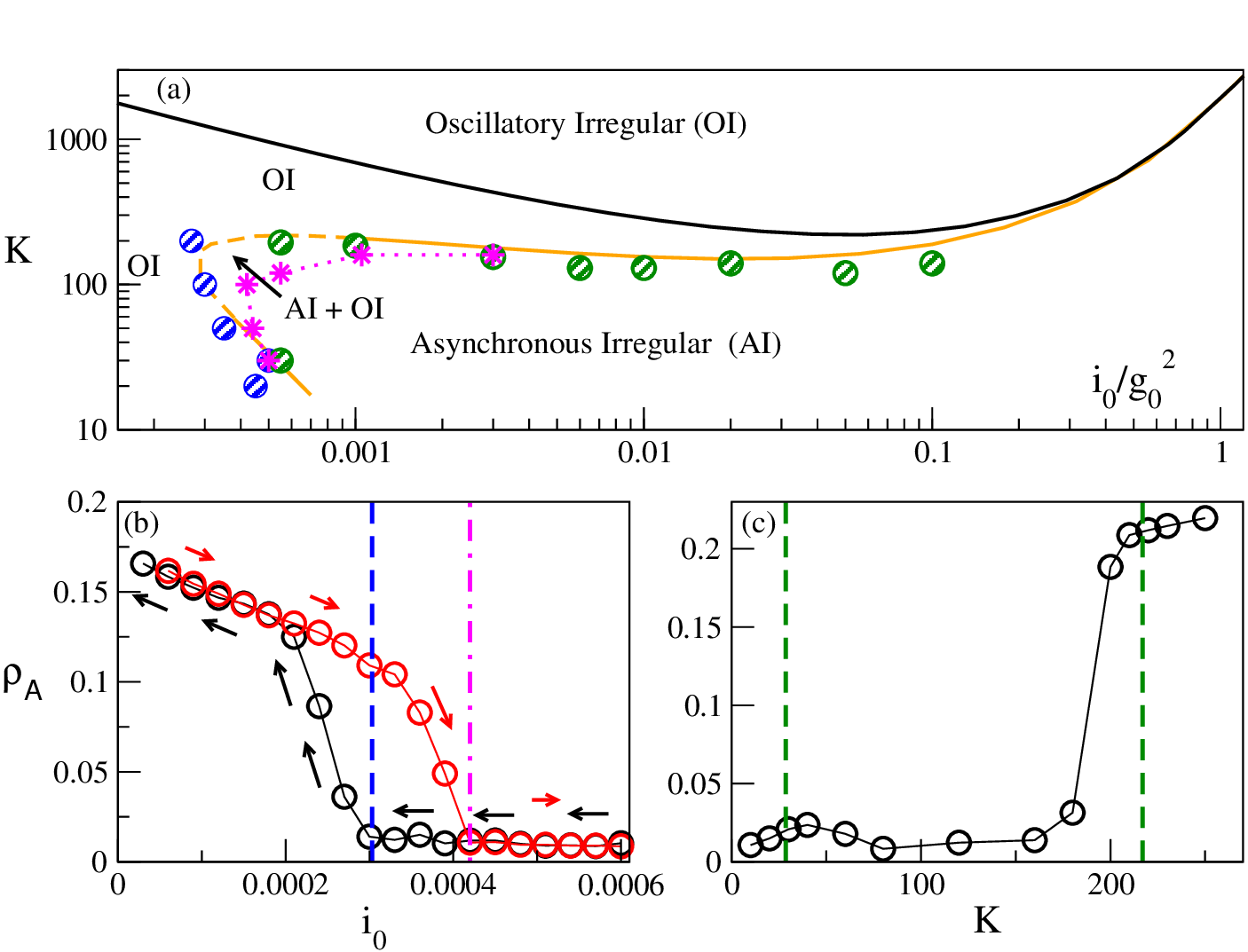}
\end{center}
\caption{(a) Phase diagram for the QIF network in the plane $(i_0/g_0^2,K)$: the black solid line is the super-critical HB line obtained within the DA; the orange solid (dashed) line is the
super- (sub-) critical HB line given by the CMF; the symbols refer to numerical
estimations of the HBs and Saddle-Node Bifurcations (SNBs). The green (blue) circles denote HBs obtained 
by performing quasi-adiabatic simulations by varying $K$ ($i_0$) for constant $i_0$ ($K$) values; the magenta stars indicate SNBs. For more details see \cite{supp}. 
(b-c) Average order parameter $ \rho_A$ versus $i_0$ ($K$) for quasi-adiabatic simulations : black circles refer to decreasing (increasing) $i_0$ ($K$), while red ones to increasing (decreasing) $i_0$ ($K$). The blue dashed line in (b) denotes the sub-critical HB given by the CMF and the magenta dot-dashed line to numerically estimated SNB; the two green dashed lines in (c) indicate the HBs given by the CMF.
The values of $\rho_A$ in panel (b) (panel (c)) refer to $K=100$ ($i_0 = 0.00055$)
averaged over 5 network realizations, with $N=80000$, for a time interval $t=30$
following a transient of 20 s. All data refer to $g_0=1$.
}
\label{fig2}
\end{figure}

The dynamics of the system (\ref{eqFP03},\ref{eqFP05})
is controlled by only two parameters: $K$ and $\alpha$. Thus, we can limit to derive a bidimensional phase diagram in the parameter plane $(K,i_0/g_0^2)$
, that will comprehensively cover the entire diversity of the macroscopic regimes observable in the network.
In particular, we have estimated the stationary solutions of Eqs. (\ref{eqFP03},\ref{eqFP05}) by truncating
the Fourier expansion in \eqref{eqFP03} to $M \ge 100$ modes in order to guarantee a numerical accuracy of ${\cal O}(10^{-12})$
for all the parameter values. The linear stability of the asynchronous state has allowed us to identify a HB line
where the oscillatory dynamics emerges: this is reported as a orange line in Fig. \ref{fig2} (a) together with the
super-critical HB line obtained within the DA (black solid line) previously reported in \cite{di2022}.
At variance with the DA the HBs induced by the shot-noise can be either super- (solid orange line) or sub-critical (dashed orange line),
thus allowing for regions where asynchronous and oscillatory regimes can coexist, see  Fig. \ref{fig2} (b).
Furthermore while for the DA the oscillatory dynamics is observable only for sufficiently large in-degree 
$K \ge K_{min} \simeq 220$, by taking into account the discrete nature of the synaptic events COs may emerge 
even  for extremely small in-degrees. Furthermore,  the asynchronous regime is always unstable 
for sufficiently small $i_0$ (large $g_0$) : namely, for for $i_0/g_0^2 < 0.00029$.
A peculiarity of the shot-noise results is that the HB line is re-entrant, thus in a certain range of 
$i_0/g_0^2$ we can have asynchronous dynamics only in a finite interval of in-degrees (as shown in Fig. \ref{fig2} (c)).

\paragraph{Network Simulations.} In order to verify the CMF predictions we have performed essentially exact numerical
simulations of the QIF network by employing a fast event-driven integration scheme \cite{tonnelier2007}, which allowed us to follow the network dynamics for long times, up to $50-100$ sec, for system of sizes $N = 10000 - 80000$ \cite{supp}. In particular, to characterize the macroscopic evolution of the network we measured the 
following indicator \cite{golomb}
\begin{equation}
\rho = \left[ \sigma^2_V /  \overline{\sigma_i^2} \right]^{1/2}
\quad {\rm where} \quad
\sigma_i^2 = \langle V_i^2 \rangle - \langle V_i \rangle^2
 \quad ,
\end{equation}
and $\sigma_V$ is the standard deviation of the mean membrane potential $\overline{V} = \sum_{i=1}^N V_i/N$,
with $\overline{\enskip \cdot \enskip}$ ($\langle \cdot \rangle$) denoting an ensemble (a time) average.
A coherent macroscopic activity is associated with a finite
value of $\rho$ (perfect synchrony corresponds to $\rho \equiv 1$),
while an asynchronous dynamics to a vanishingly small $\rho \simeq {\cal O}(1/\sqrt{N}$).
A finite size analysis of the order parameter
$\rho_A$ averaged over several different network realizations has allowed us to identify the
HBs and the Saddle-Node Bifurcations (SNBs) of limit cycles displayed in Fig. \ref{fig2}.
In particular, in Fig. \ref{fig2} (a) green (blue) circles refer to HBs identified via
quasi-adiabatic simulations by varying $K$ ($i_0$) for constant $i_0$ ($K$) values;
while the magenta stars indicate SNBs. Numerical simulations are in good agreement with
the CMF results and allowed us also the identification of a coexistence region for 
asynchronous irregular and oscillatory irregular dynamics. By irregular we mean
that the microscopic evolution is characterized by fluctuations in the instantaneous firing rates
associated to coefficient of variations \cite{cv} of ${\cal O}(1)$, as we have verified \cite{brunel1999}.
A hysteretic transition from AI to OI obtained by varying quasi-adiabatically $i_0$
is displayed in Fig. \ref{fig2} (b), the coexistence region can be clearly identified 
between the sub-critical HB (blue dashed line) and the SNB (magenta dashed line). Furthermore,
as shown in Fig. \ref{fig2} (c) for sufficiently small currents AI states are observables
only for intermediate values of the in-degrees ($K \in [50:180]$ in the considered case), bounded by regions at small ($K \le 40$) and large ($K \ge 200$) in-degrees where OI are instead present. The finite-size scaling analysis of $\rho_A$ for this specific case, revealing the different regimes, is reported in Fig. S1 in \cite{supp}.

\begin{figure}
\begin{center}
\includegraphics[width=0.95 \linewidth]{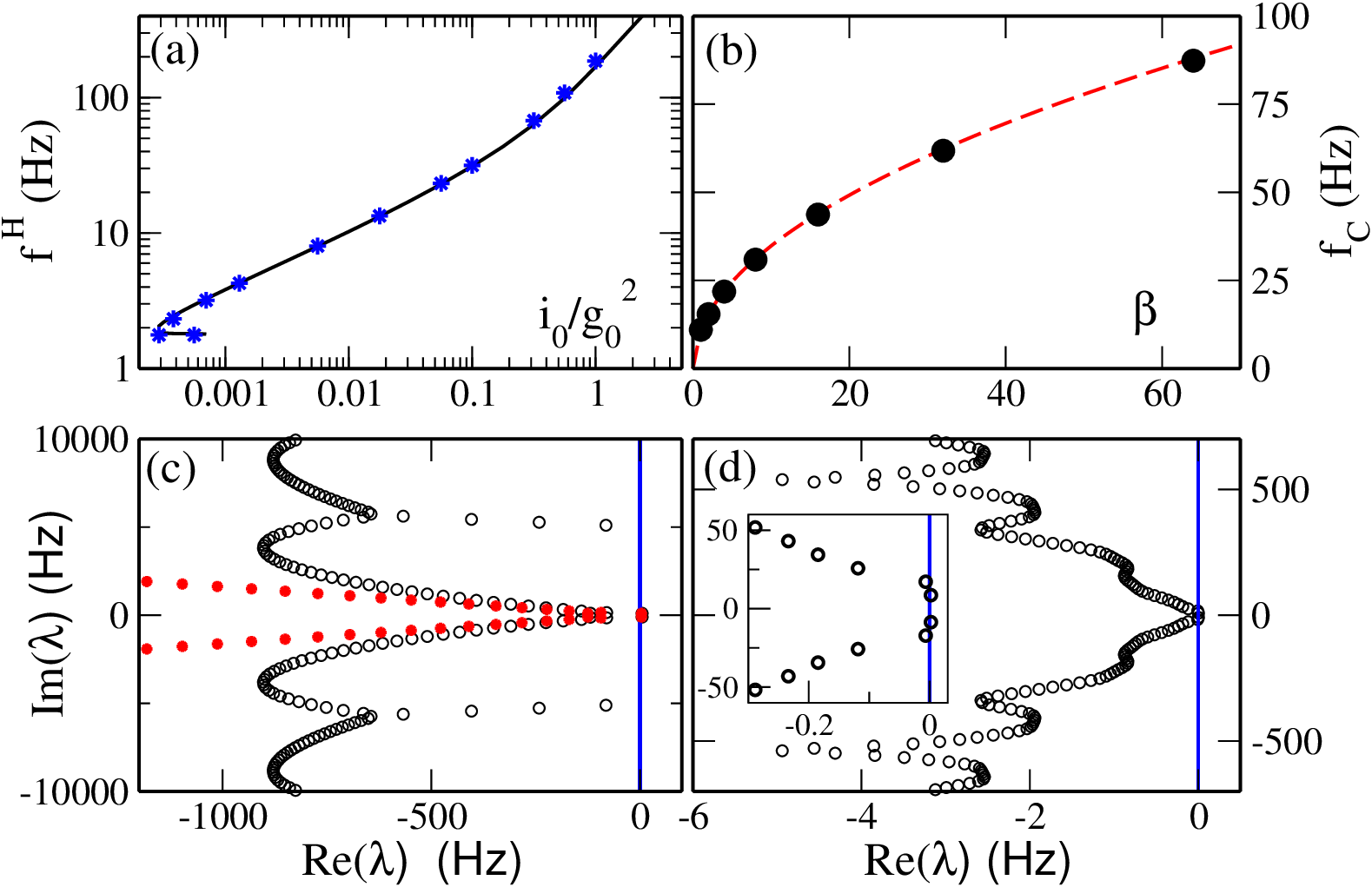}
\end{center}
\caption{(a) Frequency $f^H$ of the COs at the HB versus $i_0/g_0^2$: symbols are simulations for $N=20000$ and the solid line are the CMF results.(b) Frequency $f_{C}$ of the COs as a function of the parameter $\beta$,
where $i_0=\beta \times 0.01$, $g_0=\sqrt{\beta}$, $K=200$. Circles are network simulation data with $N=20000$ and the red dashed line
represents the curve $\nu_{CO}= 11 \sqrt{\beta}$ Hz. (c-d) Spectrum of the eigenvalues $\{\lambda_i\}$ for a stationary solution of system~(\ref{eqFP03},\ref{eqFP05}) for $(i_0/g_0^2,K)=(0.02,400)$ (c), and $(0.00055,10)$ (d). An enlargement is reported in the inset in (d). Black circles (red stars) refer to the CMF (DA). 
}
\label{fig3}
\end{figure}

At the HBs, COs emerge with
a frequency $f^H$ that is reported as a function of $i_0/g_0^2$ in Fig. \ref{fig3} (a). The comparison between the
results of the CMF approach (solid line) and of network simulations with $N=20000$ (blue stars) is very good along
the whole bifurcation line predicted by the CMF. Furthermore, $f^H$ covers a wide range
of frequencies ranging from $1.77$ Hz ($\delta$ band) to $\simeq 100$ Hz ($\gamma$ band).

As expected by the CMF analysis, the same dyanmics should be observable at fixed $K$ 
by maintaing the ratio $i_0/g_0^2$ constant. Indeed this is the case, as we have verified by considering
a state in the OI regime corresponding to $(K,i_0/g_0^2) = (200,0.01)$ and by
varying, as a function of  a control paramer $\beta$, the synaptic coupling and the current
as $g_0 = \sqrt{\beta}$  and $i_0 = \beta \times 0.01$, while $K$ stays fconstant.
We obsvered irregular dynamics characterized by an average $  \overline {CV} \simeq 0.78$ \cite{cv} and COs in the whole examined range $\beta \in [1,64]$. As expected, the only observable variation 
is in the time scale, that decreases as $1/\sqrt{I}$ \cite{supp,matteo,noi}, 
consequently the frequency $f_C$ of the COs grows proportionally to $\sqrt{\beta}$, 
thus one can observe OI dynamics induced by finite amplitude PSP in a wide frequency range by simply 
varying the parameter $\beta$ (see Fig. \ref{fig3} (a)).

\paragraph{Stability of the Asynchronous Regime: DA versus CMF.} The linearization of the 
system (\ref{eqFP03},\ref{eqFP05}) allows us to perform a linear stability analysis of the asynchronous
regime, corresponding to a constant firing rate. In particular, we have estimated
the corresponding complex spectrum $\{ \lambda_i \}$: the fixed point is
stable whenever $Re \enskip \lambda_i < 0 \quad \forall i $. Here we would like to compare the spectra obtained within the DA and the CMF to better understand the origin of the instabilities leading to oscillatory dynamics in presence of
microscopic shot-noise. As a first remark, we observe that the DA spectra are characterized besides 
the most unstable modes, which can give rise to the oscillatory instability, by modes that are strongly
damped as shown in Fig. \ref{fig3} (c). The case shown in  Fig. \ref{fig3} (c) refers to a situation
where the dynamics is well reproduced within the DA, in this case the DA eigenvalues corresponding to 
small ${\rm Im} \enskip \lambda_i$ in proximity of the Hopf instability approximate quite well the CMF spectrum.
However, while the CMF eigenvalues appear to saturate at some finite ${\rm Re} \enskip \lambda$ value, 
the DA ones do not. Despite this difference in this case the collective dynamics of the system is essentially 
controlled by the two most unstable modes, that pratically coincide within the DA and CMF approaches.

In Fig. \ref{fig3} (d) we report the CMF spectrum for a situation
where the OI regime is definitely due to the finitess of the synaptic
stimulations and not captured at all by the DA. In this case, we observe that a large part of the eigenmodes are now practically not damped, compare the scales over which  ${\rm Re} \enskip \lambda_i$ varies in Fig. \ref{fig3} (c) and (d).
Therefore, we expect that the collective dynamics is no more dominated by only the 2 most unstable modes as usually observable in the DA, but that also the marginally stable or slightly unstable modes will have a role in the coherent dynamics, see the inset of panel (d).
 
In summary, the shot-noise promotes the emergence of weakly damped eigenmodes that have a relevant role in
the instability of the asynchronous regime at sufficiently small in-degrees and that are neglected in the DA.

\paragraph{Conclusions.}

We have shown that the macroscopic phase-diagram of balanced networks is strongly influenced
by the discreteness and the finite amplitude of PSPs. In particular, we have developed
a CMF formalism by including Poissonian shot-noise which reproduces quite well the 
network simulations, at variance with the DA. 
Our mean-field analysis of the balanced state complements the previous ones, that
has been performed in the conxtext of the DA \cite{brunel1999} or in the limit $N >> K >> 1$ \cite{bal1}, and adresses 
some aspects of the neural dynamics not taken into account by the previous analysis.
A counter-intuitive aspect is the fact that COs can be observed even in extremely sparse inhibitory
networks with frequencies in a wide range from 1-2 Hz ($\delta$-band) to 100 Hz ($\gamma$-band).
Thus somehow supporting the supposition reported in  \cite{buzsaki2012} that $\gamma$-oscillations
in the hippocampus are generated by sub-networks of interneurons with low in-degrees $K \simeq 30-80$ \cite{sik1995}.

Our analysis has been limited to homogeneous networks, the inclusion of
heterogeneity in the mean-field formulation could be probably worked out by 
assuming Lorentzian distributed heterogeneities which can be
analytically integrated \cite{yakubovich1969,ott2009,montbrio2015}, 
somehow similarly to what done within the DA in \cite{noi}.
  
Quite recently, the effect of shot-noise induced by finite size fluctuations 
have been analyzed for the macrosocpic evolution of globally coupled populations 
of QIF neurons \cite{klinshov2022,klinshov2023}.
It will be interesting to combine such approach with our to fully understand the relevance 
of finite-size fluctuations for the dynamics of random sparse networks.

\begin{acknowledgments}
We acknowledge stimulating discussions with Alberto Bacci, Alberto Ferrara, Nina La Miciotta, 
Lyudmila Klimenko, Gianluigi Mongillo, Simona Olmi, Antonio Politi.
D.S.G. acknowledges the support of the CNR Short Term Mobility Programme 2021 for a visit to Istituto dei Sistemi Complessi, Sesto Fiorentino, Italy where part of this work was developed. A.T. received financial support by the Labex MME-DII (Grant No.\ ANR-11-LBX-0023-01), by CY Generations (Grant No ANR-21-EXES-0008), and together with M.V. by the ANR Project ERMUNDY (Grant No.\ ANR-18-CE37-0014) and M.V. by the Labex CORTEX (Grant No. ANR-11-LABX-0042)
of Universit\'e Claude Bernard Lyon 1 and by the the ANR via the Junior Professor Chair in Computational Neurosciences Lyon 1.
\end{acknowledgments}



\begin{thebibliography}{51}%
\makeatletter
\providecommand \@ifxundefined [1]{%
 \@ifx{#1\undefined}
}%
\providecommand \@ifnum [1]{%
 \ifnum #1\expandafter \@firstoftwo
 \else \expandafter \@secondoftwo
 \fi
}%
\providecommand \@ifx [1]{%
 \ifx #1\expandafter \@firstoftwo
 \else \expandafter \@secondoftwo
 \fi
}%
\providecommand \natexlab [1]{#1}%
\providecommand \enquote  [1]{``#1''}%
\providecommand \bibnamefont  [1]{#1}%
\providecommand \bibfnamefont [1]{#1}%
\providecommand \citenamefont [1]{#1}%
\providecommand \href@noop [0]{\@secondoftwo}%
\providecommand \href [0]{\begingroup \@sanitize@url \@href}%
\providecommand \@href[1]{\@@startlink{#1}\@@href}%
\providecommand \@@href[1]{\endgroup#1\@@endlink}%
\providecommand \@sanitize@url [0]{\catcode `\\12\catcode `\$12\catcode
  `\&12\catcode `\#12\catcode `\^12\catcode `\_12\catcode `\%12\relax}%
\providecommand \@@startlink[1]{}%
\providecommand \@@endlink[0]{}%
\providecommand \url  [0]{\begingroup\@sanitize@url \@url }%
\providecommand \@url [1]{\endgroup\@href {#1}{\urlprefix }}%
\providecommand \urlprefix  [0]{URL }%
\providecommand \Eprint [0]{\href }%
\providecommand \doibase [0]{http://dx.doi.org/}%
\providecommand \selectlanguage [0]{\@gobble}%
\providecommand \bibinfo  [0]{\@secondoftwo}%
\providecommand \bibfield  [0]{\@secondoftwo}%
\providecommand \translation [1]{[#1]}%
\providecommand \BibitemOpen [0]{}%
\providecommand \bibitemStop [0]{}%
\providecommand \bibitemNoStop [0]{.\EOS\space}%
\providecommand \EOS [0]{\spacefactor3000\relax}%
\providecommand \BibitemShut  [1]{\csname bibitem#1\endcsname}%
\let\auto@bib@innerbib\@empty
\bibitem [{\citenamefont {Schottky}(1918)}]{sc1918}%
  \BibitemOpen
  \bibfield  {author} {\bibinfo {author} {\bibfnamefont {W.}~\bibnamefont
  {Schottky}},\ }\href@noop {} {\bibfield  {journal} {\bibinfo  {journal}
  {Annalen der Physik}\ }\textbf {\bibinfo {volume} {362}},\ \bibinfo {pages}
  {541} (\bibinfo {year} {1918})}\BibitemShut {NoStop}%
\bibitem [{\citenamefont {Blanter}\ and\ \citenamefont
  {B{\"u}ttiker}(2000)}]{blanter2000}%
  \BibitemOpen
  \bibfield  {author} {\bibinfo {author} {\bibfnamefont {Y.~M.}\ \bibnamefont
  {Blanter}}\ and\ \bibinfo {author} {\bibfnamefont {M.}~\bibnamefont
  {B{\"u}ttiker}},\ }\href@noop {} {\bibfield  {journal} {\bibinfo  {journal}
  {Physics reports}\ }\textbf {\bibinfo {volume} {336}},\ \bibinfo {pages} {1}
  (\bibinfo {year} {2000})}\BibitemShut {NoStop}%
\bibitem [{\citenamefont {Lucente}\ \emph {et~al.}(2023)\citenamefont
  {Lucente}, \citenamefont {Viale}, \citenamefont {Gnoli}, \citenamefont
  {Puglisi},\ and\ \citenamefont {Vulpiani}}]{lucente2023}%
  \BibitemOpen
  \bibfield  {author} {\bibinfo {author} {\bibfnamefont {D.}~\bibnamefont
  {Lucente}}, \bibinfo {author} {\bibfnamefont {M.}~\bibnamefont {Viale}},
  \bibinfo {author} {\bibfnamefont {A.}~\bibnamefont {Gnoli}}, \bibinfo
  {author} {\bibfnamefont {A.}~\bibnamefont {Puglisi}}, \ and\ \bibinfo
  {author} {\bibfnamefont {A.}~\bibnamefont {Vulpiani}},\ }\href@noop {}
  {\bibfield  {journal} {\bibinfo  {journal} {Physical Review Letters}\
  }\textbf {\bibinfo {volume} {131}},\ \bibinfo {pages} {078201} (\bibinfo
  {year} {2023})}\BibitemShut {NoStop}%
\bibitem [{\citenamefont {Capocelli}\ and\ \citenamefont
  {Ricciardi}(1971{\natexlab{a}})}]{capocelli1971}%
  \BibitemOpen
  \bibfield  {author} {\bibinfo {author} {\bibfnamefont {R.}~\bibnamefont
  {Capocelli}}\ and\ \bibinfo {author} {\bibfnamefont {L.}~\bibnamefont
  {Ricciardi}},\ }\href@noop {} {\bibfield  {journal} {\bibinfo  {journal}
  {Kybernetik}\ }\textbf {\bibinfo {volume} {8}},\ \bibinfo {pages} {214}
  (\bibinfo {year} {1971}{\natexlab{a}})}\BibitemShut {NoStop}%
\bibitem [{\citenamefont {Tuckwell}(1988)}]{tuckwell1988}%
  \BibitemOpen
  \bibfield  {author} {\bibinfo {author} {\bibfnamefont {H.~C.}\ \bibnamefont
  {Tuckwell}},\ }\href@noop {} {\emph {\bibinfo {title} {Introduction to
  theoretical neurobiology: nonlinear and stochastic theories}}},\
  Vol.~\bibinfo {volume} {2}\ (\bibinfo  {publisher} {Cambridge University
  Press},\ \bibinfo {year} {1988})\BibitemShut {NoStop}%
\bibitem [{\citenamefont {van Vreeswijk}\ and\ \citenamefont
  {Sompolinsky}(1996)}]{bal1}%
  \BibitemOpen
  \bibfield  {author} {\bibinfo {author} {\bibfnamefont {C.}~\bibnamefont {van
  Vreeswijk}}\ and\ \bibinfo {author} {\bibfnamefont {H.}~\bibnamefont
  {Sompolinsky}},\ }\href {\doibase 10.1126/science.274.5293.1724} {\bibfield
  {journal} {\bibinfo  {journal} {Science}\ }\textbf {\bibinfo {volume}
  {274}},\ \bibinfo {pages} {1724} (\bibinfo {year} {1996})}\BibitemShut
  {NoStop}%
\bibitem [{\citenamefont {Renart}\ \emph {et~al.}(2010)\citenamefont {Renart},
  \citenamefont {de~la Rocha}, \citenamefont {Bartho}, \citenamefont
  {Hollender}, \citenamefont {Parga}, \citenamefont {Reyes},\ and\
  \citenamefont {Harris}}]{bal2}%
  \BibitemOpen
  \bibfield  {author} {\bibinfo {author} {\bibfnamefont {A.}~\bibnamefont
  {Renart}}, \bibinfo {author} {\bibfnamefont {J.}~\bibnamefont {de~la Rocha}},
  \bibinfo {author} {\bibfnamefont {P.}~\bibnamefont {Bartho}}, \bibinfo
  {author} {\bibfnamefont {L.}~\bibnamefont {Hollender}}, \bibinfo {author}
  {\bibfnamefont {N.}~\bibnamefont {Parga}}, \bibinfo {author} {\bibfnamefont
  {A.}~\bibnamefont {Reyes}}, \ and\ \bibinfo {author} {\bibfnamefont {K.~D.}\
  \bibnamefont {Harris}},\ }\href {\doibase 10.1126/science.1179850} {\bibfield
   {journal} {\bibinfo  {journal} {Science}\ }\textbf {\bibinfo {volume}
  {327}},\ \bibinfo {pages} {587} (\bibinfo {year} {2010})}\BibitemShut
  {NoStop}%
\bibitem [{\citenamefont {Brunel}(2000)}]{brunel2000}%
  \BibitemOpen
  \bibfield  {author} {\bibinfo {author} {\bibfnamefont {N.}~\bibnamefont
  {Brunel}},\ }\href {\doibase 10.1023/A:1008925309027} {\bibfield  {journal}
  {\bibinfo  {journal} {Journal of Computational Neuroscience}\ }\textbf
  {\bibinfo {volume} {8}},\ \bibinfo {pages} {183} (\bibinfo {year}
  {2000})}\BibitemShut {NoStop}%
\bibitem [{\citenamefont {Barral}\ and\ \citenamefont
  {Reyes}(2016)}]{barral2016}%
  \BibitemOpen
  \bibfield  {author} {\bibinfo {author} {\bibfnamefont {J.}~\bibnamefont
  {Barral}}\ and\ \bibinfo {author} {\bibfnamefont {A.~D.}\ \bibnamefont
  {Reyes}},\ }\href@noop {} {\bibfield  {journal} {\bibinfo  {journal} {Nature
  neuroscience}\ }\textbf {\bibinfo {volume} {19}},\ \bibinfo {pages} {1690}
  (\bibinfo {year} {2016})}\BibitemShut {NoStop}%
\bibitem [{\citenamefont {Song}\ \emph {et~al.}(2005)\citenamefont {Song},
  \citenamefont {Sj{\"o}str{\"o}m}, \citenamefont {Reigl}, \citenamefont
  {Nelson},\ and\ \citenamefont {Chklovskii}}]{song2005}%
  \BibitemOpen
  \bibfield  {author} {\bibinfo {author} {\bibfnamefont {S.}~\bibnamefont
  {Song}}, \bibinfo {author} {\bibfnamefont {P.~J.}\ \bibnamefont
  {Sj{\"o}str{\"o}m}}, \bibinfo {author} {\bibfnamefont {M.}~\bibnamefont
  {Reigl}}, \bibinfo {author} {\bibfnamefont {S.}~\bibnamefont {Nelson}}, \
  and\ \bibinfo {author} {\bibfnamefont {D.~B.}\ \bibnamefont {Chklovskii}},\
  }\href@noop {} {\bibfield  {journal} {\bibinfo  {journal} {PLoS biology}\
  }\textbf {\bibinfo {volume} {3}},\ \bibinfo {pages} {e68} (\bibinfo {year}
  {2005})}\BibitemShut {NoStop}%
\bibitem [{\citenamefont {Lefort}\ \emph {et~al.}(2009)\citenamefont {Lefort},
  \citenamefont {Tomm}, \citenamefont {Sarria},\ and\ \citenamefont
  {Petersen}}]{lefort2009}%
  \BibitemOpen
  \bibfield  {author} {\bibinfo {author} {\bibfnamefont {S.}~\bibnamefont
  {Lefort}}, \bibinfo {author} {\bibfnamefont {C.}~\bibnamefont {Tomm}},
  \bibinfo {author} {\bibfnamefont {J.-C.~F.}\ \bibnamefont {Sarria}}, \ and\
  \bibinfo {author} {\bibfnamefont {C.~C.}\ \bibnamefont {Petersen}},\ }\href
  {\doibase https://doi.org/10.1016/j.neuron.2008.12.020} {\bibfield  {journal}
  {\bibinfo  {journal} {Neuron}\ }\textbf {\bibinfo {volume} {61}},\ \bibinfo
  {pages} {301 } (\bibinfo {year} {2009})}\BibitemShut {NoStop}%
\bibitem [{\citenamefont {Miles}(1990)}]{miles1990}%
  \BibitemOpen
  \bibfield  {author} {\bibinfo {author} {\bibfnamefont {R.}~\bibnamefont
  {Miles}},\ }\href@noop {} {\bibfield  {journal} {\bibinfo  {journal} {The
  Journal of Physiology}\ }\textbf {\bibinfo {volume} {431}},\ \bibinfo {pages}
  {659} (\bibinfo {year} {1990})}\BibitemShut {NoStop}%
\bibitem [{\citenamefont {Barbour}\ \emph {et~al.}(2007)\citenamefont
  {Barbour}, \citenamefont {Brunel}, \citenamefont {Hakim},\ and\ \citenamefont
  {Nadal}}]{barbour2007}%
  \BibitemOpen
  \bibfield  {author} {\bibinfo {author} {\bibfnamefont {B.}~\bibnamefont
  {Barbour}}, \bibinfo {author} {\bibfnamefont {N.}~\bibnamefont {Brunel}},
  \bibinfo {author} {\bibfnamefont {V.}~\bibnamefont {Hakim}}, \ and\ \bibinfo
  {author} {\bibfnamefont {J.-P.}\ \bibnamefont {Nadal}},\ }\href@noop {}
  {\bibfield  {journal} {\bibinfo  {journal} {TRENDS in Neurosciences}\
  }\textbf {\bibinfo {volume} {30}},\ \bibinfo {pages} {622} (\bibinfo {year}
  {2007})}\BibitemShut {NoStop}%
\bibitem [{\citenamefont {Buzs{\'a}ki}\ and\ \citenamefont
  {Mizuseki}(2014)}]{buzsaki2014}%
  \BibitemOpen
  \bibfield  {author} {\bibinfo {author} {\bibfnamefont {G.}~\bibnamefont
  {Buzs{\'a}ki}}\ and\ \bibinfo {author} {\bibfnamefont {K.}~\bibnamefont
  {Mizuseki}},\ }\href@noop {} {\bibfield  {journal} {\bibinfo  {journal}
  {Nature Reviews Neuroscience}\ }\textbf {\bibinfo {volume} {15}},\ \bibinfo
  {pages} {264} (\bibinfo {year} {2014})}\BibitemShut {NoStop}%
\bibitem [{\citenamefont {Kisv{\'a}rday}\ \emph {et~al.}(1993)\citenamefont
  {Kisv{\'a}rday}, \citenamefont {Beaulieu},\ and\ \citenamefont
  {Eysel}}]{kisvarday1993}%
  \BibitemOpen
  \bibfield  {author} {\bibinfo {author} {\bibfnamefont {Z.~F.}\ \bibnamefont
  {Kisv{\'a}rday}}, \bibinfo {author} {\bibfnamefont {C.}~\bibnamefont
  {Beaulieu}}, \ and\ \bibinfo {author} {\bibfnamefont {U.~T.}\ \bibnamefont
  {Eysel}},\ }\href@noop {} {\bibfield  {journal} {\bibinfo  {journal} {Journal
  of comparative neurology}\ }\textbf {\bibinfo {volume} {327}},\ \bibinfo
  {pages} {398} (\bibinfo {year} {1993})}\BibitemShut {NoStop}%
\bibitem [{\citenamefont {Sik}\ \emph {et~al.}(1995)\citenamefont {Sik},
  \citenamefont {Penttonen}, \citenamefont {Ylinen},\ and\ \citenamefont
  {Buzs{\'a}ki}}]{sik1995}%
  \BibitemOpen
  \bibfield  {author} {\bibinfo {author} {\bibfnamefont {A.}~\bibnamefont
  {Sik}}, \bibinfo {author} {\bibfnamefont {M.}~\bibnamefont {Penttonen}},
  \bibinfo {author} {\bibfnamefont {A.}~\bibnamefont {Ylinen}}, \ and\ \bibinfo
  {author} {\bibfnamefont {G.}~\bibnamefont {Buzs{\'a}ki}},\ }\href@noop {}
  {\bibfield  {journal} {\bibinfo  {journal} {Journal of Neuroscience}\
  }\textbf {\bibinfo {volume} {15}},\ \bibinfo {pages} {6651} (\bibinfo {year}
  {1995})}\BibitemShut {NoStop}%
\bibitem [{\citenamefont {Buzs{\'a}ki}\ and\ \citenamefont
  {Wang}(2012)}]{buzsaki2012}%
  \BibitemOpen
  \bibfield  {author} {\bibinfo {author} {\bibfnamefont {G.}~\bibnamefont
  {Buzs{\'a}ki}}\ and\ \bibinfo {author} {\bibfnamefont {X.-J.}\ \bibnamefont
  {Wang}},\ }\href@noop {} {\bibfield  {journal} {\bibinfo  {journal} {Annual
  review of neuroscience}\ }\textbf {\bibinfo {volume} {35}},\ \bibinfo {pages}
  {203} (\bibinfo {year} {2012})}\BibitemShut {NoStop}%
\bibitem [{\citenamefont {Wildenberg}\ \emph {et~al.}(2021)\citenamefont
  {Wildenberg}, \citenamefont {Rosen}, \citenamefont {Lundell}, \citenamefont
  {Paukner}, \citenamefont {Freedman},\ and\ \citenamefont
  {Kasthuri}}]{wildenberg2021}%
  \BibitemOpen
  \bibfield  {author} {\bibinfo {author} {\bibfnamefont {G.~A.}\ \bibnamefont
  {Wildenberg}}, \bibinfo {author} {\bibfnamefont {M.~R.}\ \bibnamefont
  {Rosen}}, \bibinfo {author} {\bibfnamefont {J.}~\bibnamefont {Lundell}},
  \bibinfo {author} {\bibfnamefont {D.}~\bibnamefont {Paukner}}, \bibinfo
  {author} {\bibfnamefont {D.~J.}\ \bibnamefont {Freedman}}, \ and\ \bibinfo
  {author} {\bibfnamefont {N.}~\bibnamefont {Kasthuri}},\ }\href@noop {}
  {\bibfield  {journal} {\bibinfo  {journal} {Cell Reports}\ }\textbf {\bibinfo
  {volume} {36}} (\bibinfo {year} {2021})}\BibitemShut {NoStop}%
\bibitem [{\citenamefont {Richardson}\ and\ \citenamefont
  {Swarbrick}(2010)}]{richardson2010}%
  \BibitemOpen
  \bibfield  {author} {\bibinfo {author} {\bibfnamefont {M.~J.}\ \bibnamefont
  {Richardson}}\ and\ \bibinfo {author} {\bibfnamefont {R.}~\bibnamefont
  {Swarbrick}},\ }\href@noop {} {\bibfield  {journal} {\bibinfo  {journal}
  {Physical review letters}\ }\textbf {\bibinfo {volume} {105}},\ \bibinfo
  {pages} {178102} (\bibinfo {year} {2010})}\BibitemShut {NoStop}%
\bibitem [{\citenamefont {Iyer}\ \emph {et~al.}(2013)\citenamefont {Iyer},
  \citenamefont {Menon}, \citenamefont {Buice}, \citenamefont {Koch},\ and\
  \citenamefont {Mihalas}}]{iyer2013}%
  \BibitemOpen
  \bibfield  {author} {\bibinfo {author} {\bibfnamefont {R.}~\bibnamefont
  {Iyer}}, \bibinfo {author} {\bibfnamefont {V.}~\bibnamefont {Menon}},
  \bibinfo {author} {\bibfnamefont {M.}~\bibnamefont {Buice}}, \bibinfo
  {author} {\bibfnamefont {C.}~\bibnamefont {Koch}}, \ and\ \bibinfo {author}
  {\bibfnamefont {S.}~\bibnamefont {Mihalas}},\ }\href@noop {} {\bibfield
  {journal} {\bibinfo  {journal} {PLoS computational biology}\ }\textbf
  {\bibinfo {volume} {9}},\ \bibinfo {pages} {e1003248} (\bibinfo {year}
  {2013})}\BibitemShut {NoStop}%
\bibitem [{\citenamefont {Olmi}\ \emph {et~al.}(2017)\citenamefont {Olmi},
  \citenamefont {Angulo-Garcia}, \citenamefont {Imparato},\ and\ \citenamefont
  {Torcini}}]{olmi2017}%
  \BibitemOpen
  \bibfield  {author} {\bibinfo {author} {\bibfnamefont {S.}~\bibnamefont
  {Olmi}}, \bibinfo {author} {\bibfnamefont {D.}~\bibnamefont {Angulo-Garcia}},
  \bibinfo {author} {\bibfnamefont {A.}~\bibnamefont {Imparato}}, \ and\
  \bibinfo {author} {\bibfnamefont {A.}~\bibnamefont {Torcini}},\ }\href@noop
  {} {\bibfield  {journal} {\bibinfo  {journal} {Scientific reports}\ }\textbf
  {\bibinfo {volume} {7}},\ \bibinfo {pages} {1577} (\bibinfo {year}
  {2017})}\BibitemShut {NoStop}%
\bibitem [{\citenamefont {Droste}\ and\ \citenamefont
  {Lindner}(2017)}]{droste2017}%
  \BibitemOpen
  \bibfield  {author} {\bibinfo {author} {\bibfnamefont {F.}~\bibnamefont
  {Droste}}\ and\ \bibinfo {author} {\bibfnamefont {B.}~\bibnamefont
  {Lindner}},\ }\href@noop {} {\bibfield  {journal} {\bibinfo  {journal}
  {Journal of computational neuroscience}\ }\textbf {\bibinfo {volume} {43}},\
  \bibinfo {pages} {81} (\bibinfo {year} {2017})}\BibitemShut {NoStop}%
\bibitem [{\citenamefont {Brunel}\ and\ \citenamefont
  {Hakim}(1999)}]{brunel1999}%
  \BibitemOpen
  \bibfield  {author} {\bibinfo {author} {\bibfnamefont {N.}~\bibnamefont
  {Brunel}}\ and\ \bibinfo {author} {\bibfnamefont {V.}~\bibnamefont {Hakim}},\
  }\href@noop {} {\bibfield  {journal} {\bibinfo  {journal} {Neural
  computation}\ }\textbf {\bibinfo {volume} {11}},\ \bibinfo {pages} {1621}
  (\bibinfo {year} {1999})}\BibitemShut {NoStop}%
\bibitem [{\citenamefont {Kadmon}\ and\ \citenamefont
  {Sompolinsky}(2015)}]{bal4}%
  \BibitemOpen
  \bibfield  {author} {\bibinfo {author} {\bibfnamefont {J.}~\bibnamefont
  {Kadmon}}\ and\ \bibinfo {author} {\bibfnamefont {H.}~\bibnamefont
  {Sompolinsky}},\ }\href {\doibase 10.1103/PhysRevX.5.041030} {\bibfield
  {journal} {\bibinfo  {journal} {Phys. Rev. X}\ }\textbf {\bibinfo {volume}
  {5}},\ \bibinfo {pages} {041030} (\bibinfo {year} {2015})}\BibitemShut
  {NoStop}%
\bibitem [{\citenamefont {Monteforte}\ and\ \citenamefont {Wolf}(2010)}]{wolf}%
  \BibitemOpen
  \bibfield  {author} {\bibinfo {author} {\bibfnamefont {M.}~\bibnamefont
  {Monteforte}}\ and\ \bibinfo {author} {\bibfnamefont {F.}~\bibnamefont
  {Wolf}},\ }\href {\doibase 10.1103/PhysRevLett.105.268104} {\bibfield
  {journal} {\bibinfo  {journal} {Phys. Rev. Lett.}\ }\textbf {\bibinfo
  {volume} {105}},\ \bibinfo {pages} {268104} (\bibinfo {year}
  {2010})}\BibitemShut {NoStop}%
\bibitem [{\citenamefont {di~Volo}\ and\ \citenamefont
  {Torcini}(2018)}]{matteo}%
  \BibitemOpen
  \bibfield  {author} {\bibinfo {author} {\bibfnamefont {M.}~\bibnamefont
  {di~Volo}}\ and\ \bibinfo {author} {\bibfnamefont {A.}~\bibnamefont
  {Torcini}},\ }\href@noop {} {\bibfield  {journal} {\bibinfo  {journal} {Phys.
  Rev. Lett.}\ }\textbf {\bibinfo {volume} {121}},\ \bibinfo {pages} {128301}
  (\bibinfo {year} {2018})}\BibitemShut {NoStop}%
\bibitem [{\citenamefont {Ermentrout}\ and\ \citenamefont
  {Kopell}(1986)}]{ermentrout1986}%
  \BibitemOpen
  \bibfield  {author} {\bibinfo {author} {\bibfnamefont {G.~B.}\ \bibnamefont
  {Ermentrout}}\ and\ \bibinfo {author} {\bibfnamefont {N.}~\bibnamefont
  {Kopell}},\ }\href@noop {} {\bibfield  {journal} {\bibinfo  {journal} {SIAM
  Journal on Applied Mathematics}\ }\textbf {\bibinfo {volume} {46}},\ \bibinfo
  {pages} {233} (\bibinfo {year} {1986})}\BibitemShut {NoStop}%
\bibitem [{\citenamefont {Gutkin}(2022)}]{gutkin2022}%
  \BibitemOpen
  \bibfield  {author} {\bibinfo {author} {\bibfnamefont {B.}~\bibnamefont
  {Gutkin}},\ }in\ \href@noop {} {\emph {\bibinfo {booktitle} {Encyclopedia of
  computational neuroscience}}}\ (\bibinfo  {publisher} {Springer},\ \bibinfo
  {year} {2022})\ pp.\ \bibinfo {pages} {3412--3419}\BibitemShut {NoStop}%
\bibitem [{\citenamefont {Di~Volo}\ \emph
  {et~al.}(2022{\natexlab{a}})\citenamefont {Di~Volo}, \citenamefont {Segneri},
  \citenamefont {Goldobin}, \citenamefont {Politi},\ and\ \citenamefont
  {Torcini}}]{noi}%
  \BibitemOpen
  \bibfield  {author} {\bibinfo {author} {\bibfnamefont {M.}~\bibnamefont
  {Di~Volo}}, \bibinfo {author} {\bibfnamefont {M.}~\bibnamefont {Segneri}},
  \bibinfo {author} {\bibfnamefont {D.~S.}\ \bibnamefont {Goldobin}}, \bibinfo
  {author} {\bibfnamefont {A.}~\bibnamefont {Politi}}, \ and\ \bibinfo {author}
  {\bibfnamefont {A.}~\bibnamefont {Torcini}},\ }\href@noop {} {\bibfield
  {journal} {\bibinfo  {journal} {Chaos: An Interdisciplinary Journal of
  Nonlinear Science}\ }\textbf {\bibinfo {volume} {32}},\ \bibinfo {pages}
  {023120} (\bibinfo {year} {2022}{\natexlab{a}})}\BibitemShut {NoStop}%
\bibitem [{\citenamefont {Buzsaki}(2006)}]{buzsaki2006}%
  \BibitemOpen
  \bibfield  {author} {\bibinfo {author} {\bibfnamefont {G.}~\bibnamefont
  {Buzsaki}},\ }\href@noop {} {\emph {\bibinfo {title} {Rhythms of the
  Brain}}}\ (\bibinfo  {publisher} {Oxford University Press},\ \bibinfo {year}
  {2006})\BibitemShut {NoStop}%
\bibitem [{\citenamefont {Laing}(2018)}]{laing2018}%
  \BibitemOpen
  \bibfield  {author} {\bibinfo {author} {\bibfnamefont {C.~R.}\ \bibnamefont
  {Laing}},\ }\href@noop {} {\bibfield  {journal} {\bibinfo  {journal} {The
  Journal of Mathematical Neuroscience}\ }\textbf {\bibinfo {volume} {8}},\
  \bibinfo {pages} {1} (\bibinfo {year} {2018})}\BibitemShut {NoStop}%
\bibitem [{\citenamefont {Ratas}\ and\ \citenamefont
  {Pyragas}(2019)}]{ratas2019}%
  \BibitemOpen
  \bibfield  {author} {\bibinfo {author} {\bibfnamefont {I.}~\bibnamefont
  {Ratas}}\ and\ \bibinfo {author} {\bibfnamefont {K.}~\bibnamefont
  {Pyragas}},\ }\href@noop {} {\bibfield  {journal} {\bibinfo  {journal}
  {Physical Review E}\ }\textbf {\bibinfo {volume} {100}},\ \bibinfo {pages}
  {052211} (\bibinfo {year} {2019})}\BibitemShut {NoStop}%
\bibitem [{\citenamefont {Morris}\ and\ \citenamefont
  {Lecar}(1981)}]{morris1981}%
  \BibitemOpen
  \bibfield  {author} {\bibinfo {author} {\bibfnamefont {C.}~\bibnamefont
  {Morris}}\ and\ \bibinfo {author} {\bibfnamefont {H.}~\bibnamefont {Lecar}},\
  }\href@noop {} {\bibfield  {journal} {\bibinfo  {journal} {Biophysical
  journal}\ }\textbf {\bibinfo {volume} {35}},\ \bibinfo {pages} {193}
  (\bibinfo {year} {1981})}\BibitemShut {NoStop}%
\bibitem [{sup()}]{supp}%
  \BibitemOpen
  \href@noop {} {}\bibinfo {note} {See Supplemental Material at [URL will be
  inserted by publisher] for details on the employed neural models, on the
  integration of the neural networks as well as of the population models, an of
  the complete mean-field analysis.}\BibitemShut {Stop}%
\bibitem [{\citenamefont {Litwin-Kumar}\ and\ \citenamefont
  {Doiron}(2012)}]{bal3}%
  \BibitemOpen
  \bibfield  {author} {\bibinfo {author} {\bibfnamefont {A.}~\bibnamefont
  {Litwin-Kumar}}\ and\ \bibinfo {author} {\bibfnamefont {B.}~\bibnamefont
  {Doiron}},\ }\href {http://dx.doi.org/10.1038/nn.3220} {\bibfield  {journal}
  {\bibinfo  {journal} {Nat Neurosci}\ }\textbf {\bibinfo {volume} {15}},\
  \bibinfo {pages} {1498} (\bibinfo {year} {2012})}\BibitemShut {NoStop}%
\bibitem [{\citenamefont {Capocelli}\ and\ \citenamefont
  {Ricciardi}(1971{\natexlab{b}})}]{ricciardi}%
  \BibitemOpen
  \bibfield  {author} {\bibinfo {author} {\bibfnamefont {R.}~\bibnamefont
  {Capocelli}}\ and\ \bibinfo {author} {\bibfnamefont {L.}~\bibnamefont
  {Ricciardi}},\ }\href@noop {} {\bibfield  {journal} {\bibinfo  {journal}
  {Kybernetik}\ }\textbf {\bibinfo {volume} {8}},\ \bibinfo {pages} {214}
  (\bibinfo {year} {1971}{\natexlab{b}})}\BibitemShut {NoStop}%
\bibitem [{\citenamefont {Haskell}\ \emph {et~al.}(2001)\citenamefont
  {Haskell}, \citenamefont {Nykamp},\ and\ \citenamefont
  {Tranchina}}]{haskell2001}%
  \BibitemOpen
  \bibfield  {author} {\bibinfo {author} {\bibfnamefont {E.}~\bibnamefont
  {Haskell}}, \bibinfo {author} {\bibfnamefont {D.~Q.}\ \bibnamefont {Nykamp}},
  \ and\ \bibinfo {author} {\bibfnamefont {D.}~\bibnamefont {Tranchina}},\
  }\href@noop {} {\bibfield  {journal} {\bibinfo  {journal} {Network:
  Computation in Neural Systems}\ }\textbf {\bibinfo {volume} {12}},\ \bibinfo
  {pages} {141} (\bibinfo {year} {2001})}\BibitemShut {NoStop}%
\bibitem [{\citenamefont {Ermentrout}(2008)}]{ermentrout2008}%
  \BibitemOpen
  \bibfield  {author} {\bibinfo {author} {\bibfnamefont {B.}~\bibnamefont
  {Ermentrout}},\ }\href {\doibase 10.4249/scholarpedia.1398} {\bibfield
  {journal} {\bibinfo  {journal} {Scholarpedia}\ }\textbf {\bibinfo {volume}
  {3}},\ \bibinfo {pages} {1398} (\bibinfo {year} {2008})},\ \bibinfo {note}
  {revision \#122134}\BibitemShut {NoStop}%
\bibitem [{\citenamefont {Kralemann}\ \emph {et~al.}(2007)\citenamefont
  {Kralemann}, \citenamefont {Cimponeriu}, \citenamefont {Rosenblum},
  \citenamefont {Pikovsky},\ and\ \citenamefont {Mrowka}}]{kralemann2007}%
  \BibitemOpen
  \bibfield  {author} {\bibinfo {author} {\bibfnamefont {B.}~\bibnamefont
  {Kralemann}}, \bibinfo {author} {\bibfnamefont {L.}~\bibnamefont
  {Cimponeriu}}, \bibinfo {author} {\bibfnamefont {M.}~\bibnamefont
  {Rosenblum}}, \bibinfo {author} {\bibfnamefont {A.}~\bibnamefont {Pikovsky}},
  \ and\ \bibinfo {author} {\bibfnamefont {R.}~\bibnamefont {Mrowka}},\
  }\href@noop {} {\bibfield  {journal} {\bibinfo  {journal} {Physical Review
  E}\ }\textbf {\bibinfo {volume} {76}},\ \bibinfo {pages} {055201} (\bibinfo
  {year} {2007})}\BibitemShut {NoStop}%
\bibitem [{\citenamefont {Dolmatova}\ \emph {et~al.}(2017)\citenamefont
  {Dolmatova}, \citenamefont {Goldobin},\ and\ \citenamefont
  {Pikovsky}}]{dolmatova2017}%
  \BibitemOpen
  \bibfield  {author} {\bibinfo {author} {\bibfnamefont {A.~V.}\ \bibnamefont
  {Dolmatova}}, \bibinfo {author} {\bibfnamefont {D.~S.}\ \bibnamefont
  {Goldobin}}, \ and\ \bibinfo {author} {\bibfnamefont {A.}~\bibnamefont
  {Pikovsky}},\ }\href@noop {} {\bibfield  {journal} {\bibinfo  {journal}
  {Physical Review E}\ }\textbf {\bibinfo {volume} {96}},\ \bibinfo {pages}
  {062204} (\bibinfo {year} {2017})}\BibitemShut {NoStop}%
\bibitem [{\citenamefont {Kuramoto}(2012)}]{kuramoto2012}%
  \BibitemOpen
  \bibfield  {author} {\bibinfo {author} {\bibfnamefont {Y.}~\bibnamefont
  {Kuramoto}},\ }\href@noop {} {\emph {\bibinfo {title} {Chemical oscillations,
  waves, and turbulence}}},\ Vol.~\bibinfo {volume} {19}\ (\bibinfo
  {publisher} {Springer Science \& Business Media},\ \bibinfo {year}
  {2012})\BibitemShut {NoStop}%
\bibitem [{\citenamefont {Daido}(1992)}]{daido1992}%
  \BibitemOpen
  \bibfield  {author} {\bibinfo {author} {\bibfnamefont {H.}~\bibnamefont
  {Daido}},\ }\href@noop {} {\bibfield  {journal} {\bibinfo  {journal}
  {Progress of theoretical physics}\ }\textbf {\bibinfo {volume} {88}},\
  \bibinfo {pages} {1213} (\bibinfo {year} {1992})}\BibitemShut {NoStop}%
\bibitem [{\citenamefont {Di~Volo}\ \emph
  {et~al.}(2022{\natexlab{b}})\citenamefont {Di~Volo}, \citenamefont {Segneri},
  \citenamefont {Goldobin}, \citenamefont {Politi},\ and\ \citenamefont
  {Torcini}}]{di2022}%
  \BibitemOpen
  \bibfield  {author} {\bibinfo {author} {\bibfnamefont {M.}~\bibnamefont
  {Di~Volo}}, \bibinfo {author} {\bibfnamefont {M.}~\bibnamefont {Segneri}},
  \bibinfo {author} {\bibfnamefont {D.~S.}\ \bibnamefont {Goldobin}}, \bibinfo
  {author} {\bibfnamefont {A.}~\bibnamefont {Politi}}, \ and\ \bibinfo {author}
  {\bibfnamefont {A.}~\bibnamefont {Torcini}},\ }\href@noop {} {\bibfield
  {journal} {\bibinfo  {journal} {Chaos: An Interdisciplinary Journal of
  Nonlinear Science}\ }\textbf {\bibinfo {volume} {32}} (\bibinfo {year}
  {2022}{\natexlab{b}})}\BibitemShut {NoStop}%
\bibitem [{\citenamefont {Tonnelier}\ \emph {et~al.}(2007)\citenamefont
  {Tonnelier}, \citenamefont {Belmabrouk},\ and\ \citenamefont
  {Martinez}}]{tonnelier2007}%
  \BibitemOpen
  \bibfield  {author} {\bibinfo {author} {\bibfnamefont {A.}~\bibnamefont
  {Tonnelier}}, \bibinfo {author} {\bibfnamefont {H.}~\bibnamefont
  {Belmabrouk}}, \ and\ \bibinfo {author} {\bibfnamefont {D.}~\bibnamefont
  {Martinez}},\ }\href@noop {} {\bibfield  {journal} {\bibinfo  {journal}
  {Neural Computation}\ }\textbf {\bibinfo {volume} {19}},\ \bibinfo {pages}
  {3226} (\bibinfo {year} {2007})}\BibitemShut {NoStop}%
\bibitem [{\citenamefont {Golomb}(2007)}]{golomb}%
  \BibitemOpen
  \bibfield  {author} {\bibinfo {author} {\bibfnamefont {D.}~\bibnamefont
  {Golomb}},\ }\href {\doibase 10.4249/scholarpedia.1347} {\bibfield  {journal}
  {\bibinfo  {journal} {Scholarpedia}\ }\textbf {\bibinfo {volume} {2}},\
  \bibinfo {pages} {1347} (\bibinfo {year} {2007})}\BibitemShut {NoStop}%
\bibitem [{cv()}]{cv}%
  \BibitemOpen
  \href@noop {} {}\bibinfo {note} {The coefficient of variation $cv(i)$ for the
  neuron $i$ is the ratio between the standard deviation and the mean of the
  interspike intervals associated with its firing activity. $\overline{CV}$ is
  the ensemble average of the single neurons $cv(i)$.}\BibitemShut {Stop}%
\bibitem [{\citenamefont {Yakubovich}()}]{yakubovich1969}%
  \BibitemOpen
  \bibfield  {author} {\bibinfo {author} {\bibfnamefont {E.}~\bibnamefont
  {Yakubovich}},\ }\href@noop {} {\bibfield  {journal} {\bibinfo  {journal}
  {SOVIET PHYSICS JETP}\ }\textbf {\bibinfo {volume} {8}}}\BibitemShut
  {NoStop}%
\bibitem [{\citenamefont {Ott}\ and\ \citenamefont {Antonsen}(2009)}]{ott2009}%
  \BibitemOpen
  \bibfield  {author} {\bibinfo {author} {\bibfnamefont {E.}~\bibnamefont
  {Ott}}\ and\ \bibinfo {author} {\bibfnamefont {T.~M.}\ \bibnamefont
  {Antonsen}},\ }\href@noop {} {\bibfield  {journal} {\bibinfo  {journal}
  {Chaos: An interdisciplinary journal of nonlinear science}\ }\textbf
  {\bibinfo {volume} {19}} (\bibinfo {year} {2009})}\BibitemShut {NoStop}%
\bibitem [{\citenamefont {Montbri{\'o}}\ \emph {et~al.}(2015)\citenamefont
  {Montbri{\'o}}, \citenamefont {Paz{\'o}},\ and\ \citenamefont
  {Roxin}}]{montbrio2015}%
  \BibitemOpen
  \bibfield  {author} {\bibinfo {author} {\bibfnamefont {E.}~\bibnamefont
  {Montbri{\'o}}}, \bibinfo {author} {\bibfnamefont {D.}~\bibnamefont
  {Paz{\'o}}}, \ and\ \bibinfo {author} {\bibfnamefont {A.}~\bibnamefont
  {Roxin}},\ }\href@noop {} {\bibfield  {journal} {\bibinfo  {journal}
  {Physical Review X}\ }\textbf {\bibinfo {volume} {5}},\ \bibinfo {pages}
  {021028} (\bibinfo {year} {2015})}\BibitemShut {NoStop}%
\bibitem [{\citenamefont {Klinshov}\ and\ \citenamefont
  {Kirillov}(2022)}]{klinshov2022}%
  \BibitemOpen
  \bibfield  {author} {\bibinfo {author} {\bibfnamefont {V.~V.}\ \bibnamefont
  {Klinshov}}\ and\ \bibinfo {author} {\bibfnamefont {S.~Y.}\ \bibnamefont
  {Kirillov}},\ }\href@noop {} {\bibfield  {journal} {\bibinfo  {journal}
  {Physical Review E}\ }\textbf {\bibinfo {volume} {106}},\ \bibinfo {pages}
  {L062302} (\bibinfo {year} {2022})}\BibitemShut {NoStop}%
\bibitem [{\citenamefont {Klinshov}\ \emph {et~al.}(2023)\citenamefont
  {Klinshov}, \citenamefont {Smelov},\ and\ \citenamefont
  {Kirillov}}]{klinshov2023}%
  \BibitemOpen
  \bibfield  {author} {\bibinfo {author} {\bibfnamefont {V.}~\bibnamefont
  {Klinshov}}, \bibinfo {author} {\bibfnamefont {P.}~\bibnamefont {Smelov}}, \
  and\ \bibinfo {author} {\bibfnamefont {S.~Y.}\ \bibnamefont {Kirillov}},\
  }\href@noop {} {\bibfield  {journal} {\bibinfo  {journal} {Chaos: An
  Interdisciplinary Journal of Nonlinear Science}\ }\textbf {\bibinfo {volume}
  {33}} (\bibinfo {year} {2023})}\BibitemShut {NoStop}%
\end{thebibliography}

%

\end{document}